\documentclass[12pt,a4paper]{article}
\usepackage{jheppub,subfigure,graphicx,amsmath}
\usepackage{bbold}
\usepackage{multirow}
\usepackage[bottom]{footmisc}
\usepackage{graphicx,booktabs,array,float}
\usepackage{xcolor}
\usepackage{fmtcount}
\usepackage{caption}
\newcommand{\be}{\begin{equation}}
\newcommand{\ee}{\end{equation}} 
\newcommand{\ba}{\begin{array}}
\newcommand{\ea}{\end{array}}
\newcommand{\bea}{\begin{eqnarray}}
\newcommand{\eea}{\end{eqnarray}}
\newcommand{\nn}{\nonumber}

\newcommand{\curl}{\nabla\times}
\newcommand{\bra}[1]{\mbox{$\langle #1 |$}}
\newcommand{\ket}[1]{\mbox{$| #1 \rangle$}}
\newcommand{\abs}[1]{\left| #1 \right|}
\usepackage{tikz}
\usepackage[utf8]{inputenc}

\title{ \center  \Large Tilted Dirac cones and their topology   in \\   Holographic Materials } 
\author[a]{Jeong-Won Seo}
\author[a]{Taewon Yuk}
\author[a]{Sang-Jin Sin}
\emailAdd{1113dino@naver.com}
\emailAdd{tae1yuk@gmail.com}
\emailAdd{sjsin@hanyang.ac.kr}
\affiliation[a]{Department of Physics, Hanyang University, Seoul 04763, Korea }
\abstract{
We explore strongly correlated materials with tilted Dirac cone by introducing
a method to realize this spectral feature within a holographic setup.
Following the work by Moradpouri et al., we construct an asymptotically AdS spacetime by uplifting the vielbein of Volovik et al to tilt the flat spacetime light cone.
We then couple the resulting metric to holographic fermions and compute their spectral functions, confirming the presence of a tilted Dirac cone in momentum space.
We also calculate the topological number using the holographic Green's function and find that 
the Chern number is independent of the tilting parameter.
Additionally, we show that the optical conductivity exhibits a Drude peak even at zero chemical potential,
revealing nontrivial strong-coupling effects absent in field-theoretic models.
}

\subheader{\today }
\arxivnumber{2007.xxxxx} 
\begin{document}
	\maketitle
	
	\section{Introduction} 
	
	The exploration of electronic systems governed by the Dirac equation has unraveled a fascinating class of materials known as Dirac materials \cite{doi:10.1080/00018732.2014.927109}. These materials exhibit a distinct band structure, characterized by the  overlapping  of two Dirac cones at a single point, where 
	the Fermi surface contracts to a   point, resulting in electrons that effectively behave as massless particles with a linear dispersion relation. Such small dirac cone gives a subtle way to lead the material to be strongly interacting in the material system: while the presence the electron-hole pairs around the Fermi surface can screen any free charge inside metal very effectively, for the material with tiny Fermi surface restrict such screening mechanism, leading to the transport anomaly like breakdown of the Wiedemann–Franz law\cite{pkim} 
	and it was shown that the holographic models can effectively describe such phenomena \cite{Seo:2016vks,Seo:2017yux,Seo:2017oyh}. 
	
	Under certain circumstance,	 Dirac  materials exhibit intriguing structures that break  Lorentz symmetry  by tilting the Dirac cone \cite{katayama2006pressure, zhou2014semimetallic, goerbig2008tilted}.  Such phenomena were first observed in 
	 layered organic compound \(\alpha\text{-(BEDT-TTF)}_2I_3\) \cite{tajima2006electronic,kobayashi2007massless,tajima2009experimental,kobayashi2009theoretical,isobe2017theoretical,kajita2014molecular}. %Researchers have approached the study of such materials through two principal avenues.  
	 On the theoretical side, Volovik introduced a groundbreaking  idea,  attributing the tilt  to spacetime structure \cite{volovik2016black, volovik2021type}. This  method, although formal,   embraced the language of differential geometry and opened  an avenue to investigate spacetime structures within the context of solid-state physics \cite{farajollahpour2019solid, jafari2019electric, farajollahpour2020synthetic, weststrom2017designer}. 
	
	More recently, Moradpouri, Jafari, and Mahdi Torabian \cite{moradpouri2023holographic} uplifted  the concept of spacetime tilting in the holographic framework. They constructed an asymptotically AdS metric using  the presence of the tilted cone 
	 at the boundary flat spacetime as the boundary condition, offering a novel perspective on the behavior of strongly coupled tilted Dirac materials.  
	
	In this paper, we utilize such metric to compute the spectral function by coupling the holographic fermion to this background geometry, and demonstrated the presence of a tilted Dirac cone in momentum space. This opens the way for holography to describe a class of strongly correlated systems with tilted Dirac cones, by comparing their spectral functions with experimental data, as well as other observables derived from the band structure.
	
	We investigated the dependence on the tilting parameter and found that type-I, type-II, and type-III Dirac materials can be realized within the holographic framework. In addition to the topological number, we also calculated the optical conductivity. Notably, we find that the holographic model exhibits a finite Drude peak even at zero chemical potential, indicating the emergence of charge carriers via pair creation near the Dirac point, a feature well known in holographic system but absent in weakly coupled field-theoretic models \cite{Blake:2013bqa, Donos:2014uba}. Furthermore, the conductivity shows clear distinctions between the different tilt regimes, with enhanced spectral weight in the overtilted (type-II) case.
\vskip .3cm 
	The interplay of the topology and the interaction   is a hot theme in condensed matter. 
	Here, the topology means the  band topology defined in the momentum space. One can define a Berry potential and its curvature in terms of the eigenstate if free particle or weakly interacting quasi-particle. 
	Therefore each band structure has its own topology. 
	When interparticle interaction is strong, the band is fuzzy due to the interaction and the degree of freedom is not confined to the dispersion 
	surface $\omega=f(\bf k)$, which is a typical property of the Green function of the holographic fermions. Therefore it is questionable 
	whether the Chern number defined from the Green function indeed remained as an integer.  
	We compute the topology for the fermions with tilted Dirac cone  and  confirm that the topological number remains as integer independent of both the material type and the tilting parameter.
	\vskip .3cm 
	
	The rest of this paper is organized as follows: 	
	In Section 1, we introduce the theoretical background of tilted Dirac cones in condensed matter systems. Section 2 details the holographic model construction and demonstrates a tilted Dirac cone in the spectral function, with a tilting-independent Chern number. Section 3 explores the optical conductivity, the effect of symmetry breaking leading to different Dirac material types, and the computation of the topological number. Finally, Section 4 concludes by summarizing our findings, their relevance to real materials, and the potential for future application to Weyl fermion systems.
	
	\section{Tilted Dirac Cone}  
	
	\begin{figure}[H]
		\centering 
		\subfigure[Dirac cone]
		{\includegraphics[width=3.25cm]{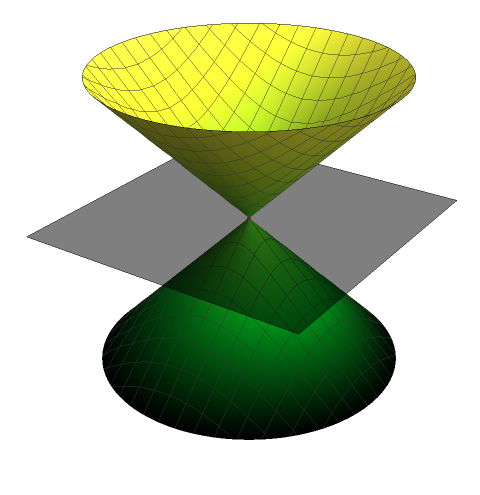}}
		\subfigure[Type \uppercase\expandafter{\romannumeral1} tilted]
		{\includegraphics[width=3.25cm]{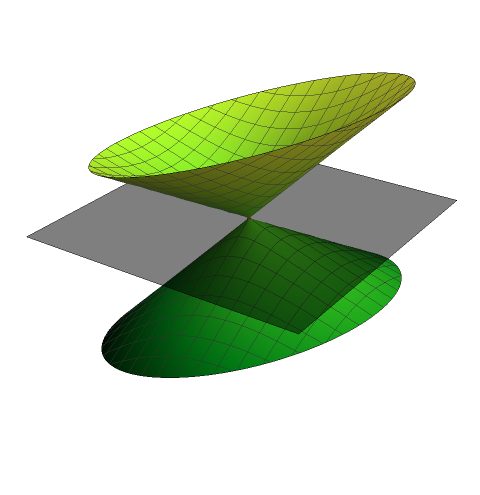}}
		\subfigure[Type \uppercase\expandafter{\romannumeral2}]
		{\includegraphics[width=3.25cm]{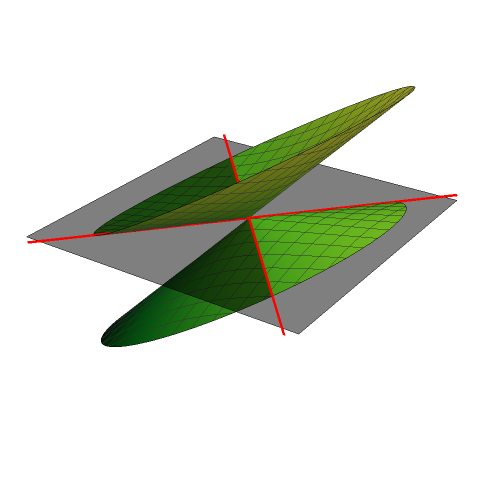}}	
		\subfigure[Type \uppercase\expandafter{\romannumeral3}]
		{\includegraphics[width=3.25cm]{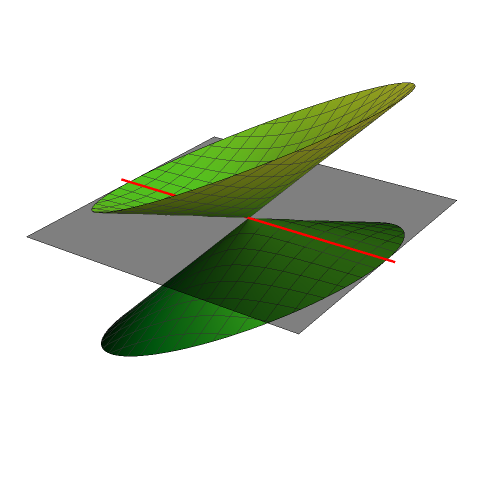}}	
		
		\subfigure[Dirac cone]
		{\includegraphics[width=3.25cm]{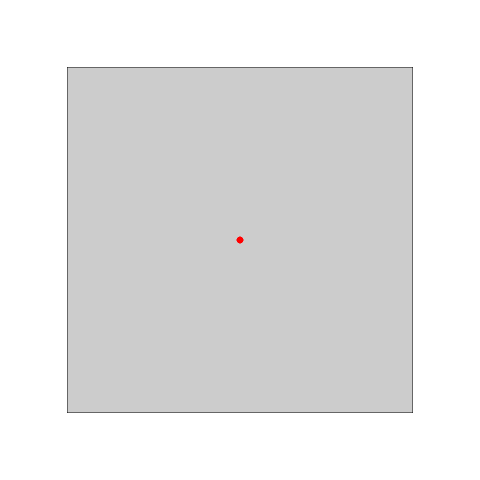}}
		\subfigure[Type \uppercase\expandafter{\romannumeral1} tilted]
		{\includegraphics[width=3.25cm]{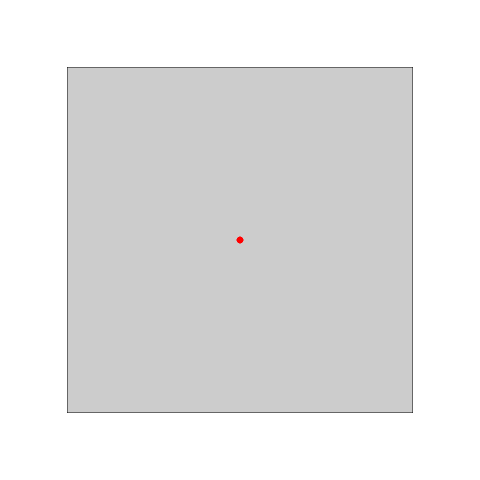}}
		\subfigure[Type \uppercase\expandafter{\romannumeral2}]
		{\includegraphics[width=3.25cm]{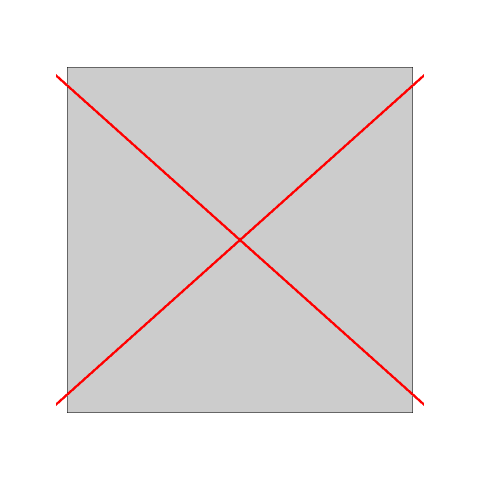}}
		\subfigure[Type \uppercase\expandafter{\romannumeral3}]
		{\includegraphics[width=3.25cm]{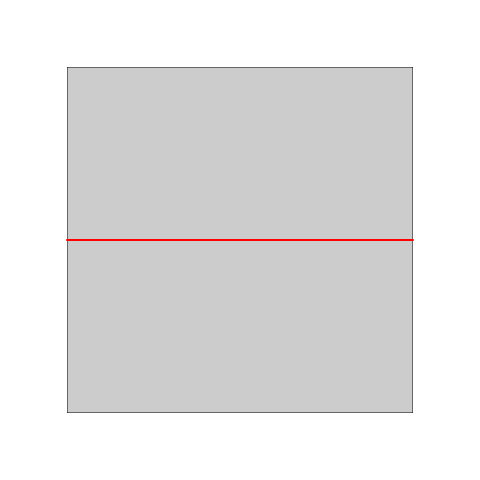}}
		\caption{\small Dirac fermion tilted along the \(\vec{\zeta}\) direction, (a,e) \(\zeta=0\), (b,f), \(\left\vert \zeta \right\vert < 1\), (c,g) \(\left\vert \zeta \right\vert > 1\), (d,h) \(\left\vert \zeta \right\vert = 1\). (a-d) Dirac cone structure of each type of Weyl fermion. (e-h) Fermi surface shape in each type of Weyl fermion. In type-I, Fermi surface is pointlike. In type-II, we have crossing nodal line. In type-III, we have one degenerate nodal line. %along \(k_\zeta\) direction
		}
		\label{Hamiltonian}
	\end{figure}
	In condensed matter system, the presence of the lattice breaks  the Lorentz symmetry of the vacuum. However, in special  cases  like graphene's honeycomb lattice \cite{katsnelson2013graphene}, an emergent Lorentz symmetry is restored locally at the low energy limit  with the speed of light replaced by the Fermi velocity, \(v_F\).
	The Hamiltonian in this limit is
	\begin{align}
		H_D=v_F\begin{pmatrix} p_z & p_x-ip_y  \\ p_x+ip_y  & -p_z \end{pmatrix}=v_F\vec{p}\cdot\vec{\sigma},
	\end{align}
	which gives a Dirac cone,  and its tip is called the  Dirac point \cite{doi:10.1080/00018732.2014.927109,novoselov2005two,zhang2005experimental}. 
	The spectrum of this system  is given in  Fig. \ref{Hamiltonian} (a,e). 
	However if a tilting is to be introduced  along the \(\vec{\zeta}\) direction, an additional term appears, \(v_F \vec{\zeta}\cdot\vec{p}\) proportional to unit matrix \(\sigma_0=\mathbb{1}\), which  breaks Lorentz symmetry. 
	In this case, the Hamiltonian near the Dirac point is :
	\begin{align}
		H=v_F\vec{p}\cdot\vec{\sigma}+v_F\vec{\zeta}\cdot\vec{p}=v_F\begin{pmatrix} p_z +\vec{\zeta}\cdot\vec{p} & p_x-ip_y  \\ p_x+ip_y  & -p_z +\vec{\zeta}\cdot\vec{p} \end{pmatrix},
	\end{align}
	with dimensionless tilting parameter \(\vec{\zeta}=(\zeta_x,\zeta_y,\zeta_z)\). 
		In this classification, \(|\zeta|<1\) corresponds to type-I and \(|\zeta|>1\) to type-II~\cite{ahn2023topological}.. 
	
	We now rewrite the Dirac equation \(i\partial_t \psi=H \psi \) in terms of vielbeins \({e}^\mu_a\) and 
	\begin{align}
		\psi=\begin{pmatrix} \psi_1 \\ \psi_2 \end{pmatrix} =\chi e^{-ik_{\mu}x^{\mu}}. 
	\end{align}
	as follows :
	\begin{align}
		{e}^\mu_a \sigma^a p_\mu \psi=0, \quad  or \quad 
		{e}^\mu_a \sigma^a k_\mu \chi=0, 
	\end{align}
	for \(\sigma^a=(\mathbb{1},\vec{\sigma})\), \(p_\mu=i\partial_\mu\), \(x_\mu=(t,\vec{x})\), \(k_\mu=(-\omega,\vec{k})\).
	From this, we can read off the vielbein fields of Dirac fermion \cite{volovik2021type} :
	\begin{align}
		{e}^\mu_a =\begin{pmatrix} 1 & \zeta_x & \zeta_y & \zeta_z \\ 0 & 1 & 0 & 0\\ 0 & 0 & 1 & 0 \\ 0 & 0 & 0 & 1 \end{pmatrix}
	\end{align} 
	These vielbein fields formally mimic  the effective gravity experienced by the tilted Dirac fermions inflat spacetime.
	
	\section{Holographic Model}
	\subsection{Flat Space Metric for a Tilted cone}
	We first introduce the effective flat space metric from the  above vielbein, 
	\begin{align}
		g^{\mu\nu}={e}^\mu_a{e}^\nu_b\eta^{ab}=\begin{pmatrix} -1 & \zeta_j \\ \zeta_j & \delta_{ij}-\zeta_i\zeta_j \end{pmatrix}
	\end{align} 
	This effective metric can be identified with the acoustic (inverse) metric, commonly observed in moving liquids and superfluids, where the fluid velocity acts as the shift velocity \cite{volovik2021type}, though we will not pursue this analogy here.
	\begin{align}
		g_{\mu\nu}=\begin{pmatrix} -1+\zeta^2 & \zeta_j \\ \zeta_j & \delta_{ij} \end{pmatrix}, & \quad \text{With} \quad \zeta^2=\sum_i\zeta_i^2
	\end{align}
	In \(\vec\zeta=0\) case, \(g_{\mu\nu}=\eta_{\mu\nu}\) , the  Minkowski space-time. In general, this metric resembles a Galilean boost. It is noteworthy that, for Dirac materials with upright cones, the emergent spacetime is Minkowski. However, introducing tilt breaks Lorentz symmetry and deforms the Minkowski structure, leading to deviations from it. For simplicity we set  \(\zeta_y=0\), \(\zeta_z=0\). 
	
	We now want to construct a metric which is asymptotically AdS and have above metric as its 
	boundary  behavior  with proper scaling behavior in the holographic coordinate. Namely 
	we seek a metric $ g_{\mu\nu}^{Bulk}$ which satisfies the Einstein equation and 
	obeys the boundary condition 
	\[ \lim_{u\to 0}  u^2g_{\mu\nu}^{Bulk}\sim g_{\mu\nu}^{boundary}  \quad for \quad \mu, \nu \neq u\]
	with 
	\begin{align}
		g^{boundary}_{\mu\nu}=\begin{pmatrix} -1+\zeta_x^2 & \zeta_x & 0 & 0 \\ \zeta_x & 1 & 0 & 0 \\0 & 0 & 1 & 0 \\0 & 0 & 0 & 1 \end{pmatrix} . 
	\end{align}
	Such a metric describes the effective gravity with tilted Dirac cone.  
	With the ansatz  \cite{moradpouri2023holographic} 
	\begin{align}
		g_{\mu\nu}^{Bulk}=\frac{1}{u^2}\begin{pmatrix} (-1+\zeta_x^2)f(u) & \zeta_x f(u) & 0 & 0 & 0\\ \zeta_x f(u) & \frac{1-\zeta_x^2f(u)}{1-\zeta_x^2} & 0 & 0 & 0 \\0 & 0 & 1 & 0 & 0 \\0 & 0 & 0 & 1 & 0 \\ 0 & 0 & 0 & 0 & \frac{1}{f(u)} \end{pmatrix}.
	\end{align}
	the unknown function \(f(u)\) can be determined from the Einstein equation, yielding
	\[ f(u)=1-\left( \frac{u}{u_H}\right)^4. \]
	In standard form, the metric can be written as 
	\begin{align*}
		ds^2=&\frac{1}{u^2}\frac{du^2}{f(u)}+\frac{1}{u^2}\left[ -\left( 1-\zeta_x^2\right) f(u)dt^2+2\zeta_xf(u)dtdx+\frac{1-\zeta_x^2f(u)}{1-\zeta_x^2}dx^2+dy^2+dz^2 \right].  
	\end{align*}

	\subsection{Fermion Action}
	The  action for the  holographic fermion   is given by the sum $S=S_{g}+S_{\psi}+S_{bdy}$ with	
	\begin{align} 
		&S_{ g} =\frac{1}{2\kappa}  \int^{\beta}_0 d\tau \int_{u=\epsilon} d\mathbf{x}\int^{\epsilon}_{u_h}du\left[\sqrt{-g}\left( R+\Lambda \right) -\sqrt{-\gamma}K+\sqrt{-\gamma}\left( \frac{3}{L}+\frac{L}{4}R\right) \right]  \\
		& S_{ \psi} =i\int d^{5}x\; \sqrt{-g}\bar{\psi}\left[ E_{\underline{a}}^{\mu} \Gamma^{\underline{a}}\left( \partial_\mu+\frac{1}{4}\omega_{\underline{\nu\lambda},\mu}\Gamma^{\underline{\nu\lambda}}\right) -m\right]  \psi \\
		& S_{ bdy}=\frac{i}{2} \int_{bdy} d^{4}x\;\sqrt{-h} {\bar\psi}\psi\label{bdryaction},
	\end{align}
	where metric \(g\) is  \(g^{Bulk}\) and  $$\kappa=8G_N, \quad \gamma_{ij}=\frac{L^2}{\epsilon}g_{ij}(\epsilon, x), \quad n^u=L^{-1}u\sqrt{f(u)}, \quad K=n^u \partial_u \sqrt{\gamma}/{\sqrt\gamma}, $$ \cite{moradpouri2023holographic} and $\omega_{\nu\lambda,\mu}$ is the spin connection with bulk vielbein fields witch is defined by 
	$$g^{\mu\nu}=E_{a}^{\mu}E_{b}^{\nu}\eta^{ab}, $$ 
	with boundary condition 
	\[\lim_{u\to 0} \frac{1}{u}E_{a}^{\mu}= e_{a}^{\mu}  \quad for \; \mu, a \neq u. \]
	The bulk vielbein can be written explicitly as
	\begin{align*} 
		E_{a}^{\mu}= u  
		\begin{pmatrix} -\frac{\sqrt{1-\zeta_x^2f(u)}}{\sqrt{\left( 1-\zeta_x^2\right)f(u) }} & \zeta_x\frac{\sqrt{\left( 1-\zeta_x^2\right)f(u) }}{\sqrt{1-\zeta_x^2f(u)}} & 0 & 0 & 0 \\ 0 & 
			\frac{\sqrt{1-\zeta_x^2f(u) }}{\sqrt{1-\zeta_x^2}} & 0 & 0 & 0  \\
			0 & 0 & 1 & 0 & 0 \\
			0 & 0 & 0 & 1 & 0 \\
			0 & 0 & 0 & 0 & \sqrt{f(u)} \end{pmatrix}.
	\end{align*} 
	The gamma matrices are defined as
\begin{align}
	\Gamma^{\underline{t}}&=\sigma_1 \otimes i \sigma_2, \; \Gamma^{\underline{x}}=\sigma_1 \otimes \sigma_1,\;
	\Gamma^{\underline{y}}=\sigma_1 \otimes \sigma_3, \;\Gamma^{\underline{z}}=\sigma_2 \otimes \sigma_1, \;\Gamma^{\underline{u}}=\sigma_3 \otimes \sigma_0,
\end{align}
	and geometry is specified by 
	\begin{align}
		h= {g}{{g}^{uu}}, \quad f(u)=1-\left( \frac{u}{u_H}\right)^4, \quad u_H=\frac{1}{\pi T} \sqrt{1 - \zeta_x^2},
		\label{eq:geo}
	\end{align}
	{\it Dirac equations } are given by   
	\be
	\begin{aligned}
		E_{\underline{a}}^{\mu} \Gamma^{\underline{a}}\left( \partial_\mu+\frac{1}{4}\omega_{\underline{\nu\lambda},\mu}\Gamma^{\underline{\nu\lambda}}\right) \psi-m \psi=0
	\end{aligned}
	\ee
	Substituting the Dirac field ansatz :
	\be
	\label{wave1}
	\psi(u)=(-gg^{uu})^{-1/4}e^{-ik_{\mu}x^{\mu}}\chi(u).
	\ee
	the resulting equations are
	\bea  \label{eom} 
		\Big[  \Gamma^{\underline{u}}\partial_u  -i\frac{\sqrt{F(u)}}{f(u)}\omega \Gamma^{\underline{t}}&& -i\zeta_x\frac{1}{\sqrt{F(u)}}k_x  \Gamma^{\underline{t}}-i\sqrt{\frac{F(u)}{f(u)}}k_x\Gamma^{\underline{x}} \nn \\
		&&-\frac{i}{\sqrt{f(u)}}\left( k_y\Gamma^{\underline{y}}+k_z\Gamma^{\underline{z}}+\frac{m}{u}\right) \Big] \chi =0, 
	\eea 
	where $ F(u)=\frac{1-\zeta_x^2f(u)}{(1-\zeta_x^2)} $ is the backreaction factor 
	 due to the tilt, ensuring the asymptotic AdS condition. Notice that near the boundary     $ \lim_{u\rightarrow0} F(u)\sim1$. 
	 %$$ \lim_{u\rightarrow0} F(u)=\lim_{u\rightarrow0}\frac{1-\zeta_x^2f(u)}{(1-\zeta_x^2)}\sim1, $$ 
The way tilting parameter couples to the fermion is  the same as that of the symmetric tensor coupling \(h_{ti}\) 
that we previously introduced \cite{byun2025symmetrictensorcouplingholographic}. Also that 
for  \(\zeta_x=0\) case, we have \(F(u)\mid_{\zeta_x=0}=1\), and we can get back to the usual Dirac equation in AdS. 
%as we known 
%	\begin{align}
%		\left[ \Gamma^{\underline{u}}\partial_u-\frac{i}{f(u)}\omega \Gamma^{\underline{t}}-\frac{i}{\sqrt{f(u)}}\left( k_x\Gamma^{\underline{x}}+k_y\Gamma^{\underline{y}}+k_z\Gamma^{\underline{z}}+\frac{m}{u}\right) \right] &\chi =0.
%	\end{align}
	
\subsection{Green’s Function}
\label{sec:greens}

We decompose the bulk Dirac spinor according to the ansatz
\eqref{wave1}.  Writing
$$
\psi=(\psi_{+},\psi_{-})^{\mathsf T}
$$
and factoring out the plane wave, we obtain
$$
\chi=(\chi_{+},\chi_{-})^{\mathsf T},
$$
By varying the bulk action\(S_{bulk}=S_g+S_{\psi}\), with respect to $\psi$ and adding the variation of the \(S_{bdy}\), one finds that the total action variation can be expressed solely in terms of $\chi_{+}$ when the equations of motion are satisfied \cite{Iqbal:2009fd}.
\begin{equation}
	\begin{gathered}
		S_{bdy} = -i\int_{\partial\mathcal{M}}d^{4}x\bar{\chi}\chi  =  -\int_{\partial\mathcal{M}}d^{4}x\,\chi^{\dagger}_{+}\chi_{-} + \mathrm{h.c.}
	\end{gathered}
\end{equation}
Therefore, we interpret the boundary quantities of $\chi_{+}$ and $\chi_{-}$ as  $\xi^{(S)}$ and $\xi^{(C)}$, corresponding to the two-component source and condensation at the boundary respectively.
Substituting this decomposition into the bulk Dirac equation \eqref{eom}, we obtain two coupled
first–order equations
\begin{align}
	\partial_{u}\xi^{(S)}+\mathbb{M}_{1}\,\xi^{(S)}
	+\mathbb{M}_{2}\,\xi^{(C)} &=0, \label{eq:eomS}\\
	\partial_{u}\xi^{(C)}+\mathbb{M}_{3}\,\xi^{(C)}
	+\mathbb{M}_{4}\,\xi^{(S)} &=0, \label{eq:eomC}
\end{align}
with
\begin{align*}
	\mathbb{M}_{1}&=-\mathbb{M}_{3}
	=-\frac{i\,m}{u\sqrt{f(u)}}\,\sigma_{0}, \\[4pt]
	\mathbb{M}_{2}&=-\mathbb{M}_{4}
	=\frac{i}{\sqrt{f(u)}}\!
	\begin{pmatrix}
		k_{y}-ik_{z} &
		\sqrt{\tfrac{F(u)}{f(u)}}\,\omega
		+\dfrac{1+\zeta_x\sqrt{f(u)}}{\sqrt{F(u)}}\,k_{x}\\[2pt]
		-\sqrt{\tfrac{F(u)}{f(u)}}\,\omega
		+\dfrac{1+\zeta_x\sqrt{f(u)}}{\sqrt{F(u)}}\,k_{x} &
		k_{y}+ik_{z}
	\end{pmatrix},
\end{align*}
where $\sigma_{0}$ is the identity matrix.

For $|m|<\tfrac12$ the independent solutions scale near the conformal
boundary ($u\to0$) as
$$
\xi^{(S)}\simeq u^{m}\,\mathcal{J}, \quad \xi^{(C)}\simeq u^{-m}\,\mathcal{C},
$$
with two component, $u$–independent spinors
$\mathcal{J}$ (source) and $\mathcal{C}$ (condensate).
Evaluating the on–shell variation of the total action shows that
$(\chi_{1},\chi_{2})\equiv\xi^{(S)}$ indeed furnishes the canonical
Dirichlet data.  The boundary action becomes
\[
S_{\mathrm{bdy}}
=\frac12\!\int d^{4}x\,
\bigl(-\mathcal{J}^{\dagger}\sigma_{2}\mathcal{C}+\text{h.c.}\bigr).
\]
Thus the matrix–valued retarded Green’s function is obtained from
linear response as
\begin{equation}
	G_{R}(\omega,\mathbf{k})
	=-\sigma_{2}\,\mathbb{C}_{0}\,\mathbb{S}_{0}^{-1},
\end{equation}
where $\mathbb{S}_{0}$ and $\mathbb{C}_{0}$ denote the leading
boundary coefficients of $\xi^{(S)}$ and $\xi^{(C)}$, respectively.
Equivalently, we introduce the bulk matrix
\(
\mathbb{G}(u)=\mathbb{C}(u)\mathbb{S}^{-1}(u),
\)
Eqs.~\eqref{eq:eomS}–\eqref{eq:eomC} combine into a Riccati flow
equation whose solution yields
\begin{equation}
	G_{R}(\omega,\mathbf{k})
	=\lim_{u\to0}u^{2m}\,\mathbb{G}(u).
\end{equation}

In the pure AdS limit ($T\to0$, so $f(u)\to1$), the flow equation
admits a closed‐form solution, giving
\begin{equation}
	\label{AnGreen}
	G_{R}=\frac{1}
	{\sqrt{k_{x}^{\,2}+k_{y}^{\,2}+k_{z}^{\,2}
			-\bigl(\omega+\zeta_{x}k_{x}\bigr)^{2}}}\!
	\begin{pmatrix}
		\omega+\zeta_{x}k_{x}+k_{z} & k_{x}-ik_{y}\\[2pt]
		k_{x}+ik_{y} & \omega+\zeta_{x}k_{x}-k_{z}
	\end{pmatrix}
\end{equation}
which smoothly reproduces the standard massless
$\mathrm{AdS}_{5}$ result when $\zeta_x=0$.
	
	\subsection{Spectral Function}
\begin{figure}[H]
		\centering 
		\subfigure[\(\zeta_x=0\)]
		{\includegraphics[width=3.25cm]{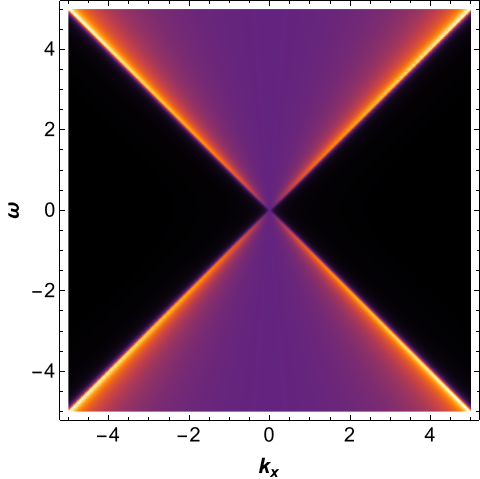}}
		\subfigure[\(\zeta_x=0.5\)]
		{\includegraphics[width=3.25cm]{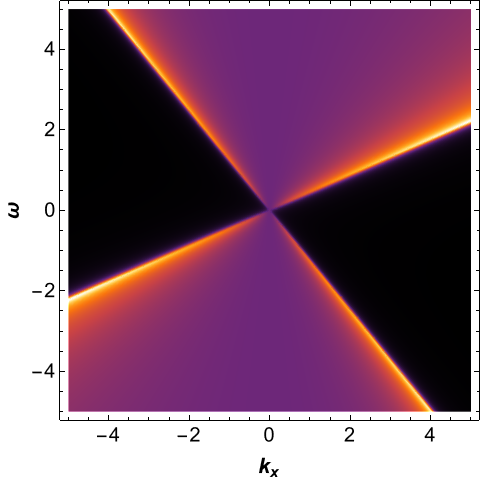}}
		\subfigure[\(\zeta_x=1.5\)]
		{\includegraphics[width=3.25cm]{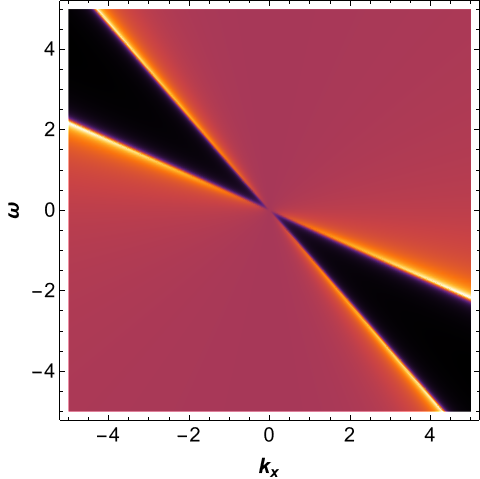}}
		\subfigure[\(\zeta_x\simeq1\)]
		{\includegraphics[width=3.25cm]{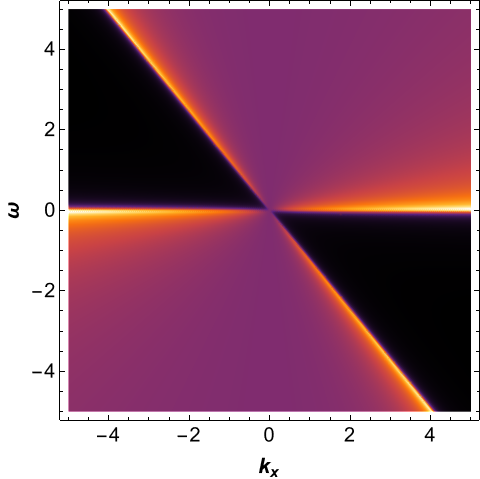}}
		
		\subfigure[\(\zeta_x=0\)]
		{\includegraphics[width=3.25cm]{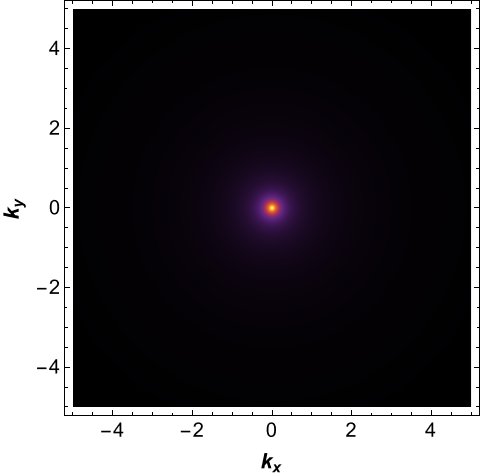}}
		\subfigure[\(\zeta_x=0.5\)]
		{\includegraphics[width=3.25cm]{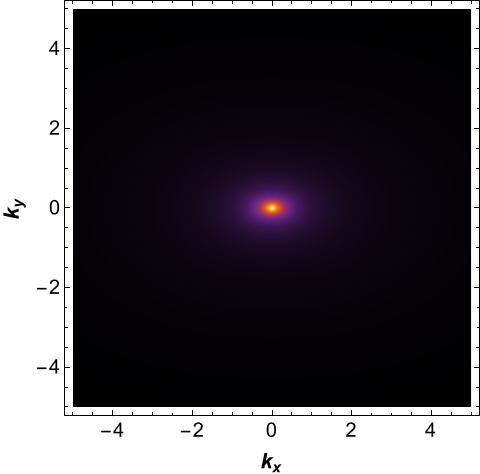}}
		\subfigure[\(\zeta_x=1.5\)]
		{\includegraphics[width=3.25cm]{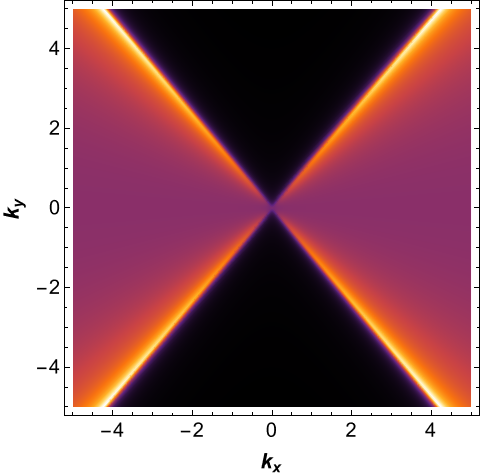}}
		\subfigure[\(\zeta_x\simeq1\)]
		{\includegraphics[width=3.25cm]{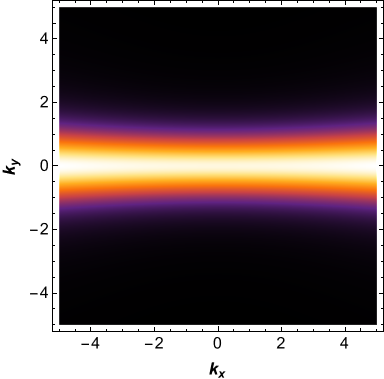}}
		\caption{\small  Holographic Spectral function of  tilted Dirac fermion with \(T=0.01\). 
			Figures in the upper line are in $(\omega,k_x)$ plane while those in the lower line are in $( k_x,k_y)$ plane. 
			(a,e) Normal Dirac cone in \(\zeta_x=0\) (b,f) Type-I tilted Dirac cone  for  \(\zeta_x<1\). (c,g)   Type-II   for \(\zeta_x>1\) with over tilted Dirac cone. (c,g) type-III for    \(\zeta_x\simeq1\), which is  the critical case.}
		\label{SF}
	\end{figure}

	thus the spectral density of holographic fermion is given by 
	\begin{align}
		A(k, w) =Im\left[ Tr\left[ \mathbb{G}(k, w)\right]\right]=\frac{2 (w+\zeta_x k_x)}{\sqrt{k_x^2+k_y^2+k_z^2-(w+\zeta_x k_x)^2}}
	\end{align} 
	For \(\zeta_x=0\), the spectral density coincides with that of the standard Dirac cone \cite{oh2021ginzberg}. Since the cone is the locus of the zero of the denominator,  it is evident that the tilting parameter \(\zeta_x\) induces a  rotation  in 
	\( (\omega,k_x)\) plane with the tilting angle 
	\[ \tan\theta = \zeta_x , \]
	highlighting the   origin of the tilting phenomenon.
	We  plotw holographic spectral function (SF)  in the Fig. \ref{SF}.
	In Fig. \ref{SF} (a,e), we plot the fermion SF for various values of \(\zeta_x\). Fig. \ref{SF} (b,f),with \(\zeta_x=0\),   Corresponds to the original dirac cone. Fig. \ref{SF} (b-d, g-h) \(\zeta_x\neq0\)  is for the tilted dirac cone.  The broken symmetry in the holographic boundary metric also influences the fermion spectral function. 
	\subsection{Comparison with Experimental Data}
\begin{figure}[H] 
		\centering
		\includegraphics[width=12cm]{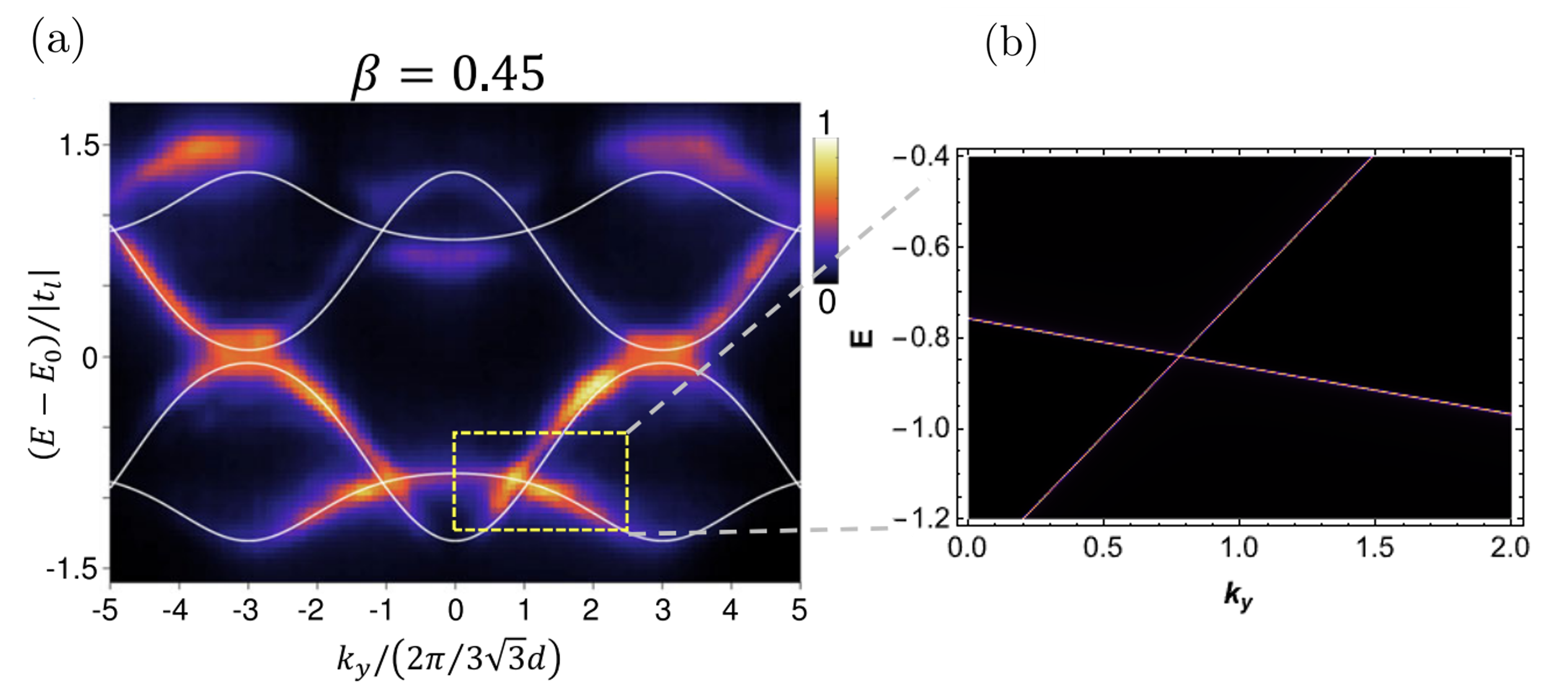}
		\caption{\small   holographic spectrum vs experimental data. (a) Photoluminescence  intensity \cite{milicevic2019type}  in photonic orbital graphene.   (b) spectrum of holographic  theory. The  Dirac cone in the yellow box in figure (a) is compared with figure (b).}	\label{comfig}
	\end{figure}

			Figure \ref{comfig} (a), shows the photoluminescence (PL) intensity data  of 'photonic orbital graphene' (POG). In the experiment of Ref. \cite{milicevic2019type}, the tilting was induced by controlling the carbon–carbon distance \(d'\), which in turn determines the hopping parameter \(t_L\) in the \(x\)-direction of the honeycomb lattice.  The authors of \cite{milicevic2019type} introduced the parameter \(\beta\)   as the ratio \(\beta \equiv t'_L/t_L\) between the two  hopping parameters, where \(t'_L\) denotes  hopping in  other direction \cite{milicevic2019type}. For \(\beta\sim1\), a flat band structure is observed. The emergence of a flat band often signals a strongly correlated system, which can be effectively studied within holography. Therefore, comparison of the experimental results in this region with holographic calculations is a meaningful approach.
			
				Fig. \ref{comfig} (a) illustrates the PL intensity of POG  at \(\beta=0.45\). The parameter \(\beta\) is given by  \cite{milicevic2019type} 
			$${v_{y }}=\frac{3\beta}{4}\sqrt{\frac{1-\beta^2}{1+\beta^2}}, \quad v_{F}=\frac{\sqrt{3}}{4}\sqrt{1-\beta^2}, $$
		which allows one to determine the tilting parameter \(\zeta_y\) in terms of \(\beta\): 
		$$ \zeta_y=\frac{{v_{y }}}{v_{F}}=\frac{\sqrt{3}\beta}{\sqrt{1+\beta^2}}.$$

	For the case \(\beta=0.45\), one obtains \(\zeta_y\sim0.71\). Fig. \ref{comfig} (b) shows the fermion spectrum of a tilted Dirac cone in the  holographic model with tilting parameter \(\zeta_y\sim0.71\). % derived from \(\beta=0.45\).  
	For fermion mass \(m=1/2\), proper scale and the shifting parameter \(B_x\), the spectrum takes the depicted form. 
	
	By linear fitting of the experimental data in figure \ref{comfig},  (a) we can determine the slopes \(-0.12\) and \(0.97\) \cite{milicevic2019type}. From these slopes we determine tilting parameter \(\zeta_y\sim0.78\). This computed value closely matches the holographically obtained value.
	This example demonstrates how  holographic model  quantitatively  describes the tilted Dirac cone system  by fitting the experimental data.

		\begin{figure}[H]
		\centering 
		\subfigure[\(m=0\)]
		{\includegraphics[width=4.5cm]{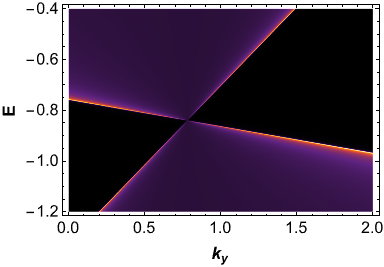}}
		\subfigure[\(m=\left| 1/8\right| \)]
		{\includegraphics[width=4.5cm]{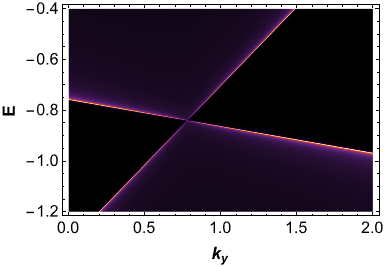}}
		\subfigure[\(m=\left| 1/2\right|\)]
		{\includegraphics[width=4.5cm]{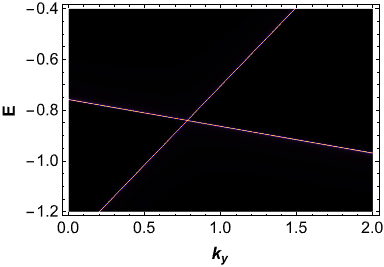}}
		\caption{\small  Holographic Spectral function of  tilted Dirac fermion various value of Holographic fermion mass \(m\).} \label{figm}
	\end{figure}
		As shown in Fig.\ref{figm}, by tuning the holographic fermion mass within the model, we can also explore strong correlation effects. As the fermion mass increases, the spectrum becomes sharper and more closely resembles the experimental results.

	\subsection{Topological Number}
	We now turn to the band topology. One of the strong interaction effects is to render the band structure fuzzy. Since band topology was originally defined for non-interacting fermions, it is natural to ask whether the topology of the band structure changes or remains in the presence of strong interactions. There are two primary methods for calculating the topological number: the eigenvector method and the Green’s function method \cite{PhysRevLett.113.136402, wang2012simplified}.
	
	{\it Green’s function method :}
	Using the analytic Green's function given in Eq.~(\ref{AnGreen}), we can express the Berry curvature of the holographic tilted Dirac material using the general formula in terms of the Berry curvature in terms of the exact Green's function :
	\begin{align}
		\mathcal{F}_c &= \left(\nabla \times \mathcal{A} \right)_c \\
		&= \frac{1}{3!} \int_{-\infty}^{\infty} \frac{d\omega}{2\pi} \text{Tr} \left[ \epsilon_{\mu \nu \rho c} \mathbb{G} \left( \partial_{\mu}\mathbb{G}^{-1} \right) \mathbb{G} \left( \partial_{\nu}\mathbb{G}^{-1} \right) \mathbb{G} \left( \partial_{\rho}\mathbb{G}^{-1} \right) \right].
	\end{align}
	Here \( \mathcal{A} \) denotes the Berry connection and \( \mathcal{F}_c \) is the generalized Berry flux. For the analytic Green's function in Eq.~(\ref{AnGreen}), the trace evaluates explicitly to:
	\begin{align}
		\text{Tr} \left[ \epsilon_{\mu \nu \rho c} \mathbb{G} \left( \partial_{\mu}\mathbb{G}^{-1} \right) \mathbb{G} \left( \partial_{\nu}\mathbb{G}^{-1} \right) \mathbb{G} \left( \partial_{\rho}\mathbb{G}^{-1} \right) \right] = -\frac{12 (1 - \zeta_x^2)^{3/2}}{\left( (1 - \zeta_x^2)\vec{k}^2 + (\omega + \zeta_x k_x)^2 \right)^2} k_c,
	\end{align}
	where \( \vec{k}^2 = k_x^2 + k_y^2 + k_z^2 \). Using this, the Berry flux becomes
	\begin{align}
		\mathcal{F}_i&=-\frac{1}{3!}\int_{-\infty}^{\infty}\frac{d\omega}{2\pi}\frac{12(-1+\zeta_x^2)^{\frac{3}{2}}}{\left((-1+\zeta_x^2)\vec{k}^2+(\omega+\zeta_xk_x)^2 \right)^{2}}k_i \\ &=-\lim_{L\rightarrow\infty}\frac{k_i}{2\pi\vec{k}^{\frac{3}{2} }}\left[  \frac{\omega+\zeta_xk_x}{  (-1+\zeta_x^2)\vec{k}^{2}+(\omega+\zeta_xk_x)^2 }\sqrt{\frac{-1+\zeta_x^2}{\vec{k}}}+\tan^{-1}\left(\frac{w+\zeta_x k_x}{\sqrt{(-1+\zeta_x^2)\vec{k}^2}}\right)\right]^{L}_{-L} \nonumber \\
		&=-\frac{k_i}{2\left(k^2_x+k^2_y+k^2_z \right)^{\frac{3}{2} }}.	 
	\end{align}
	Note that the result is independent of $\zeta_{x}$ showing the semi-universal  character of the Berry curvature, whose integration gives a true topological number.  \\
	
	{\it Eigenvector Method :}
	An alternative derivation employs the topological Hamiltonian method, based on the zero-frequency Green's function:
	\begin{align}
		\mathcal{H}_t(\vec{k}) &= -\mathbb{G}^{-1}(0, \vec{k}) \\
		&= -\frac{1}{\sqrt{(1 - \zeta_x^2) k_x^2 + k_y^2 + k_z^2}} 
		\begin{pmatrix}
			\zeta_x k_x + k_z & k_x - i k_y \\
			k_x + i k_y & \zeta_x k_x - k_z
		\end{pmatrix}.
	\end{align}
	The eigenvalues of this matrix are given by
	\begin{align}
		\epsilon_1 &= \frac{\zeta_x k_x - |\vec{k}|}{\sqrt{(1 - \zeta_x^2) k_x^2 + k_y^2 + k_z^2}}, \\
		\epsilon_2 &= -\frac{\zeta_x k_x - |\vec{k}|}{\sqrt{(1 - \zeta_x^2) k_x^2 + k_y^2 + k_z^2}}.
	\end{align}
	The corresponding normalized eigenvectors \( \ket{n_1}, \ket{n_2} \) are independent of \( \zeta_x \) :
	\begin{align*}
		\ket{n_1}= \left( \frac{\sqrt{k_x^2+k_y^2}\left( k_z-\abs{\vec{k}}\right) }{\sqrt{2}(k_x+ik_y)\sqrt{\vec{k}^2}-k_z\abs{\vec{k}}}, \ \frac{\sqrt{k_x^2+k_y^2} }{\sqrt{2}\sqrt{\vec{k}^2}-k_z\abs{\vec{k}}} \right) \\
		\ket{n_2}= \left( \frac{\sqrt{k_x^2+k_y^2}\left( k_z+\abs{\vec{k}}\right) }{\sqrt{2}(k_x+ik_y)\sqrt{\vec{k}^2}+k_z\abs{\vec{k}}}, \ \frac{\sqrt{k_x^2+k_y^2} }{\sqrt{2}\sqrt{\vec{k}^2}+k_z\abs{\vec{k}}} \right). 
	\end{align*}
	This implies that the Berry curvature is also independent of \( \zeta_x \). Since the Berry flux is defined by
	\begin{align*}
		\mathcal{F}_i&= \curl \bra{n} \partial \ket{n}
	\end{align*}
	Using eigen vector above We can calculate Berry flux 
	\begin{align*}
		\mathcal{F}_i&= \curl \bra{n} \partial \ket{n} \\
		&= -\frac{k_i}{2\left(k^2_x+k^2_y+k^2_z \right)^{\frac{3}{2} }}. 
	\end{align*}
Again, the independence  of the curvature  on tilting angle automatically guarantees the 
preservation of 
  the topological invariant under the smooth deformation by tilting.
	\begin{figure}[H] 
		\centering
		\includegraphics[width=5cm]{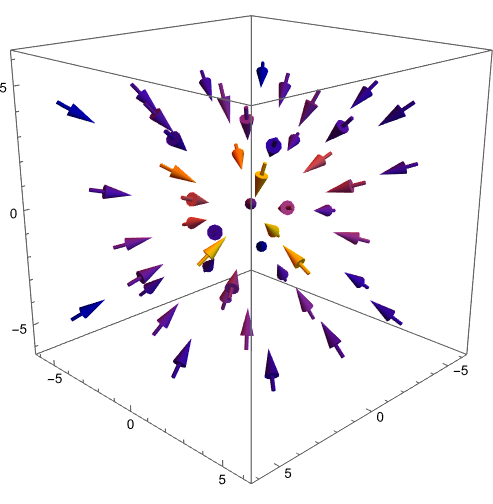}
		\caption{Three-dimensional vector plot of the Berry curvature \( \mathcal{F} \) of a tilted Dirac cone. Vectors converge at the Dirac point.} \label{curv}
	\end{figure}
%	Therefore 
%	both the Green’s function and eigenvector methods yield the same result for the Berry flux. 
%	 :
%	\begin{align} 
%		\mathcal{F}_i = -\frac{k_i}{2 \left(k_x^2 + k_y^2 + k_z^2 \right)^{\frac{3}{2} }}. 
%	\end{align}
	This confirms that the Berry flux is independent of the tilting parameter \(\zeta_x\). Fig.\ref{curv} presents a three-dimensional vector plot of the Berry curvature, illustrating the flux distribution of the tilted Dirac cone. The vectors converge at the Dirac point, demonstrating the topological nature of the system.  It is interesting to notice that the flux field does not change at all under the tilting deformation. Usually the external strain modifies the dynamics of the electrons but our result predicts that such strain will not affect the anomalous Hall effect  induced by the Berry potential.  
	
	The  Chern number  can be calculated as 
	\begin{align}
		C_n&=\frac{1}{2\pi}\oint \mathcal{F}_c \cdot dS= -1
	\end{align}
	Note that the Chern number is indepent of \(\zeta_x\),  showing the shape independence of the 
	topological number.
		
	\section{Tilted Dirac Cone as the Holographic Background Metric}
	The metric is backreacted by the presence of  the tilting source and this metric is coupled to the U(1) gauge field as well as the  fermionic fluctuations. Therefore it is interesting to   explore other  consequence  of the tilting in holographic theory. 
	\subsection{Conductivity}
	
We now investigate the optical conductivity of the boundary theory dual to a tilted AdS$_{d+2}$ black brane. This background geometry is holographically dual to a strongly coupled quantum field theory with emergent gravity corresponding to tilted Dirac cone structure. The effective metric is chosen to encode the anisotropic causal structure associated with a tilted dispersion relation, in analogy with condensed matter systems discussed in \cite{Volovik2003, Moradpouri:2021nxi}.

\begin{figure}[H] 
	\centering
	\subfigure[Type \uppercase\expandafter{\romannumeral1} : \(\zeta_x<1\)]
	{\includegraphics[width=6cm]{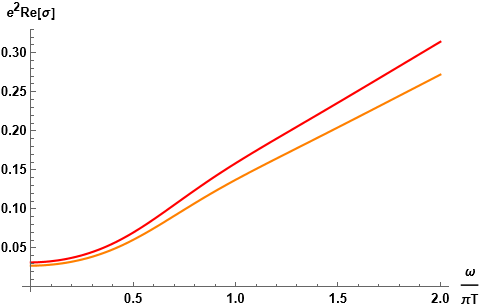}} 
	\hskip 1cm 
	\subfigure[Type \uppercase\expandafter{\romannumeral2} : \(\zeta_x>1\)]
	{\includegraphics[width=6cm]{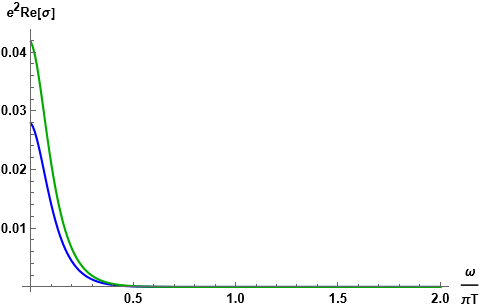}}
	\caption{Optical conductivity in tilted Dirac cone backgrounds.  The red curve corresponds to the untilted case ($\zeta_x=0$).  Orange, blue and green curves show the effect of finite tilt parameters $\zeta_x=0.5$, $1.5$ and $1.8$, respectively.}
	\label{cond}
\end{figure}

The bulk Maxwell dynamics are governed by the action
\begin{equation}
	S = -\frac{1}{4 g_{5}^2} \int d^{5}x \sqrt{-g} F_{ab} F^{ab},
\end{equation}
We consider a gauge fluctuation in the $x$-direction and $t$-direction of the form
\begin{equation}
	A_x(u,t) = a_x(u) e^{-i\omega t}
\end{equation}
and derive its equation of motion from the Maxwell action. The linearized equation of motion becomes
\begin{align}
	&a_x'' - \left( \frac{1}{u} - \frac{f'}{f} \right) a_x' + \frac{\omega^2}{(1 - \zeta_x^2) f^2} a_x = 0, 
\end{align}
Varying the on–shell Maxwell action yields the boundary current
\begin{align}
	&J^x(u)=-\frac{1}{g_{5}^{2}}\,
	\sqrt{-g}\,F^{ux}(u)=-\frac{1}{g_{5}^{2}}\,
	\sqrt{-g}\,g^{uu}g^{xx}\partial_{u}A_x(u).
	\label{Jx}
\end{align}
For the tilted background, with
$\sqrt{-g}=u^{-5}$, $g^{xx}=u^{2}(1-\zeta_x^{2})$, $g^{uu}=u^{2}f$ and $g_{5}^{2}\equiv e^{2}$, we obtain
\begin{align}
	&J^x(u)
	=-\frac{1}{e^{2}}\,
	\left( 1-\zeta_x^{2}\right)u^{-1} f(u)\,
	a_x'(u).
\end{align}
The electric field is defined as
\begin{equation}
	E_x=-\partial_t A_x=-i\omega a_x(u), 
\end{equation}
Combining \(J^x\) with $E_x$, one obtains the optical conductivity
\begin{equation}
	\sigma_{xx}(\omega)=
	\frac{1}{e^{2}}\,
	\frac{1-\zeta_x^{2}}{i\omega}\;
	\lim_{u\to0}
	\frac{1}{u}\,
	\frac{a_x'(u)}{a_x(u)}
\end{equation}
Figure~\ref{cond} displays the numerically evaluated optical conductivity in both (a) type-I ($\zeta_x<1$) and (b) type-II ($\zeta_x>1$) regimes of AdS$_{5}$ black brane.

In the untilted Dirac material (red line in Fig.~\ref{cond} (a)) the conductivity increases linearly with the frequency $\omega$  as reported in anisotropic holographic models \cite{Kovtun:2008kx,Cremonini:2017qwq} and in field-theoretic approaches \cite{hou2023effects}. Introducing a finite but subcritical tilt ($0<\zeta_x<1$) suppresses the conductivity slightly, yielding smaller values than in the untilted case. By contrast, in the overtilted (type-II) phase [Fig.~\ref{cond} (b)] the linear scaling breaks down and a pronounced Drude-like structure dominates.  
We observe that both the Drude peak height and weight grow monotonically with the tilt parameter.
This behavior is expected because the tilt enlarges the Fermi surface around the Dirac point for the type II, changing  the small Fermi surface  to a large one so that it becomes more typical metallic behavior. 
It is rather surprising that the holographic computation reproduces such metalic behavior as well as the anomalous small fermi-surface case. 	

In contrast to the suppression observed for the type-I tilted Dirac cone, the type-II regime exhibits an overall enhancement of conductivity as the tilt increases. this is duo to strong-coupling effects near the Dirac point at zero temperature. In particular, at $\mu=0$ the Drude contribution vanishes in weakly interacting models, making it difficult to quantify its influence. By contrast, in the strongly coupled holographic framework—even at $\mu=0$—pair creation near the Dirac point of massless Dirac materials produces a finite density of charge carriers, giving rise to a nonvanishing Drude peak. 
	
\subsection{Effect of Symmetry Breaking on Tilting Background Metric}
\begin{figure}[H]
	\centering 
	\subfigure[\(\zeta_x=0\)]
	{\includegraphics[width=4cm]{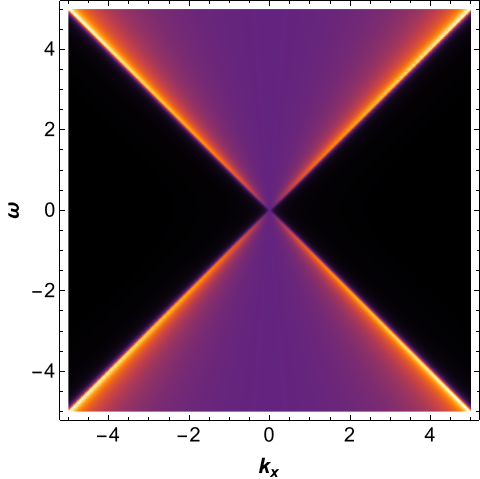}}
	\subfigure[\(\zeta_x=0.5\)]
	{\includegraphics[width=4cm]{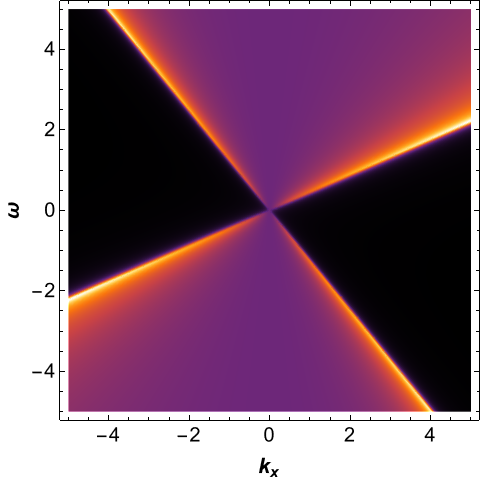}}
	\subfigure[\(\zeta_x=1.5\)]
	{\includegraphics[width=4cm]{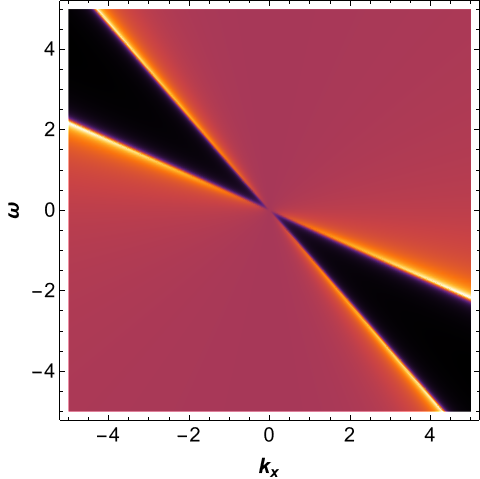}}
	\caption{Analytic spectral function of a non-interacting holographic tilted Dirac fermion}
\end{figure}
We can give symmetry breaking by fermion interaction to tilted Dirac fermion by bulk interaction term $$\int_{bulk}\Phi_{ \mu\nu...} {\bar\psi}\left( E_{\underline{a}}^{\mu}E_{\underline{b}}^{\nu}\cdots\Gamma^{\underline{ab...}}\right) \psi,$$ According to \cite{oh2021ginzberg}, such Yukawa coupling impart special features to Dirac cone depending on what symmetry is broken, and it gives same features to tilted Dirac cone.
\subsubsection{Gapping}
\begin{figure}[H]
	\centering 
	\subfigure[\(\zeta_x=0\), \(M=1\)]
	{\includegraphics[width=4cm]{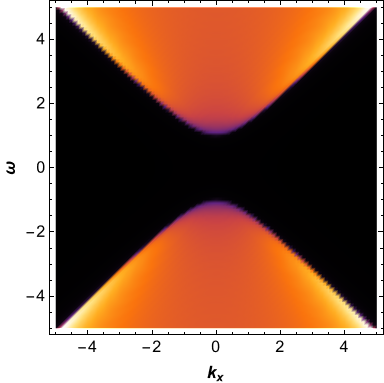}}
	\subfigure[\(\zeta_x=0.5\), \(M=1\)]
	{\includegraphics[width=4cm]{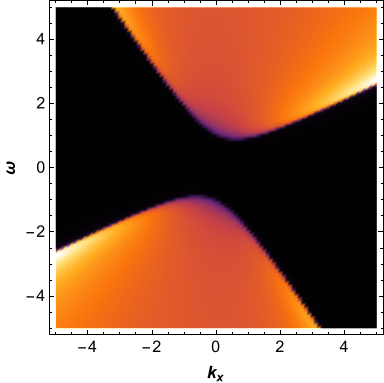}}
	\subfigure[\(\zeta_x=1.5\), \(M=1\)]
	{\includegraphics[width=4cm]{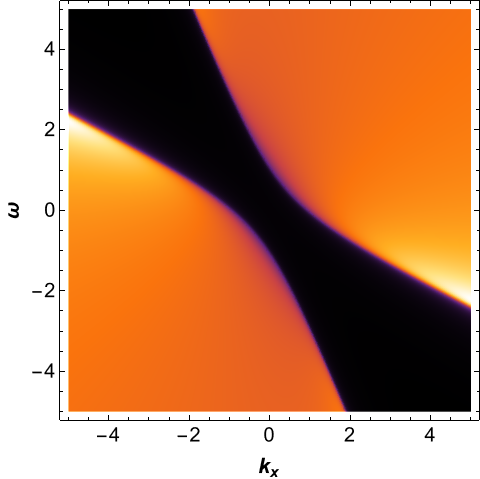}}
	\caption{Analytic spectral function of holographic tilted Dirac fermion with scalar type bulk interaction term \(\Phi(u)=M u\) for \(M>0\)}
\end{figure}
\subsubsection{Flat Band}
we can also observe zero modes (e.g., flat band) in the spectrum through some symmetry breaking \cite{oh2021ginzberg}. Hoever tilting of Dirac fermion does not affect on this zero mode, it only tilt dirac fermion only. For example the bulk term $$\int_{bulk}B_{xy} {\bar\psi}\left( E_{\underline{a}}^{x}E_{\underline{b}}^{y}\Gamma^{\underline{ab}}\right) \psi$$ gives flat band structure to fermion, but tilting of Dirac fermion does not change flat band structure :
\begin{figure}[H]
	\centering 
	\subfigure[\(\zeta_x=0\), \(B_{xy}=1\)]
	{\includegraphics[width=4cm]{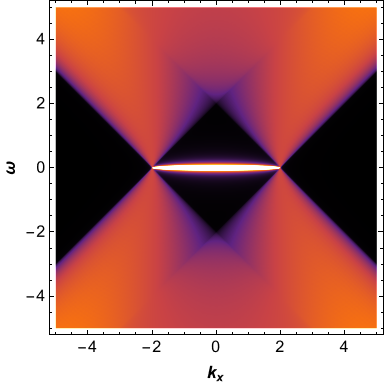}}
	\subfigure[\(\zeta_x=0.5\), \(B_{xy}=1\)]
	{\includegraphics[width=4cm]{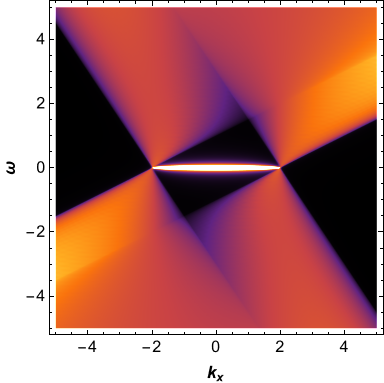}}
	\caption{Analytic spectral function of holographic tilted dirac fermion with tensor type bulk interaction term \(B_{xy}\).}
\end{figure}
\subsubsection{Topological Liquid}
\begin{figure}[H]
	\centering 
	\subfigure[\(\zeta_x=0\), \(M=-1\)]
	{\includegraphics[width=4cm]{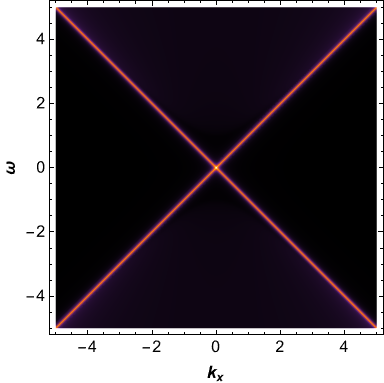}}
	\subfigure[\(\zeta_x=0.5\), \(M=-1\)]
	{\includegraphics[width=4cm]{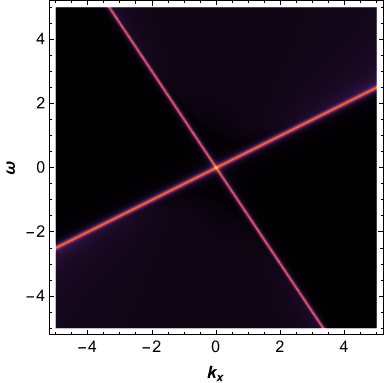}}
	\caption{Analytic spectral function of holographic tilted dirac fermion with scalar type bulk interaction term \(\Phi(u)=M u\) for \(M<0\) }
\end{figure}
When the sign of scalar type interaction parameter is negative, a crossed pole appears, which is formed along the surface of the Dirac cone. With a tilting metric at the boundary, this pole tilts in the same direction as the Dirac cone.
	
	\section{Conclusion}
	In this paper, we explored the phenomenon of tilted Dirac cones in holographic framework. 
	 Starting from  Volovik's  tilting metric of flat spacetime,  and  embedding  it   into the holographic bulk, one can construct a bulk metric which encodes the tilting effect.
We then coupled the metric to the  holographic fermions and calculated the spectral function and  the conductivity.% as well as the topological number of the band structure. 
	 We also coupled the tilted holographic fermions to the bulk order parameters and investigated their effect on the tilted  fermion.
	 This approach allowed us to model and realize tilted Dirac cones  in the  holographic setup. 
	 It would be interesting to compare our method with the holographic mean field theory developed recently
	\cite{oh2021ginzberg}.
	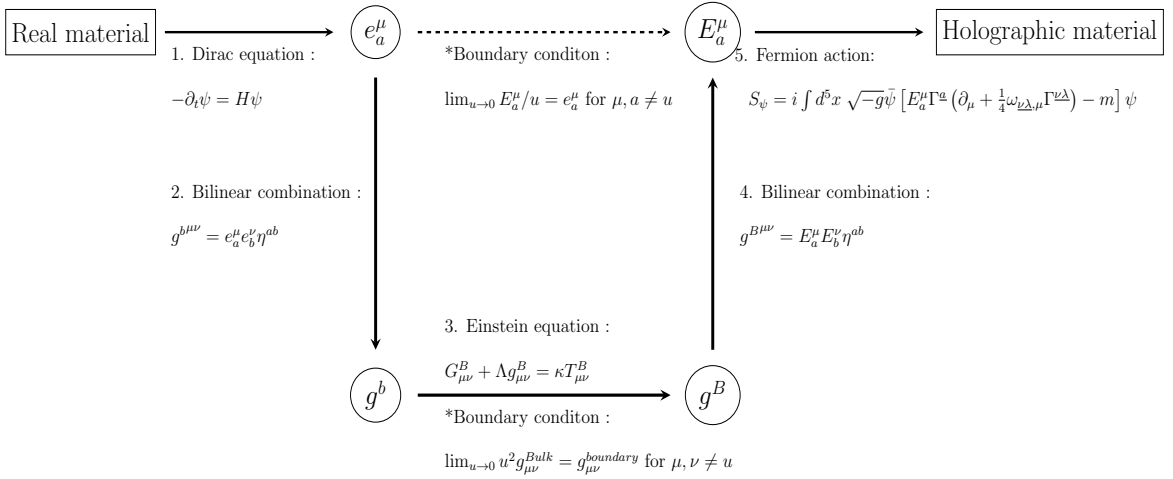
\begin{figure}[H]
		\centering
		\resizebox{6 in}{2.5 in}{%
			\begin{tikzpicture}[scale=1]
				\tikzstyle{style1} = [circle,draw,fill=white,minimum size=10mm,font=\Large];
				\tikzstyle{style2} = [rectangle,draw,fill=white,minimum size=10mm,font=\Large];
				\tikzstyle{style3} = [rectangle,draw,minimum size=10mm,font=\Large];
				\node (node0) at (-11 ,4) [style2] {Real material};
				\node (node5) at (12,4) [style2] {Holographic material};
				\draw [-stealth,line width=1.5pt](5,4) -- (9,4);
				\node[right] at (-9,3.5) {1. Dirac equation :};
				\node[right] at (-9,2.5) { \(-\partial_t \psi=H \psi\)};
				\draw [-stealth,line width=1.5pt](-9,4) --(-5,4);
				\node[left] at (8,3.5) {5. Fermion action:};
				\node[right] at (4.7,2.5) { \(S_{ \psi} =i\int d^{5}x\; \sqrt{-g}\bar{\psi}\left[ E_{a}^{\mu} \Gamma^{\underline{a}}\left( \partial_\mu+\frac{1}{4}\omega_{\underline{\nu\lambda},\mu}\Gamma^{\underline{\nu\lambda}}\right) -m\right]  \psi\)};
				\draw [-stealth,line width=2pt](4,-3) -- (4,3);
				\node[right] at (-2.5,3.5) {*Boundary conditon :};
				\node[right] at (-2.5,2.5) { \(\lim_{u\to 0}  E_{a}^{\mu}/u=e_{a}^{\mu} \) for \(\mu, a \neq u\)};
				\draw [stealth-,line width=2pt](-4,-3) -- (-4,3);
				\draw [-stealth,line width=2pt](-3,-4) -- (3,-4);
				\draw [-stealth,line width=1.5pt,dash pattern=on 3pt off 3pt](-3,4) -- (3,4);
				%	\node(dots) at (-3,-4.5){\vdots};
				\node (node1) at (4,4) [style1] {\(E_{a}^{\mu}\)};
				\node (node2) at (-4,4) [style1] {\(e_{a}^{\mu}\)};
				\node (node3) at (4,-4) [style1] {\(g^{B}\)};
				\node (node4) at (-4,-4) [style1] {\(g^{b}\)};
				\node[right] at (-2.5,-2.5) {3. Einstein equation :};
				\node[right] at (-2.5,-3.5) { \(G^B_{\mu\nu}+\Lambda g^B_{\mu\nu}=\kappa T^B_{\mu\nu} \)};
				\node[right] at (-2.5,-4.5) {*Boundary conditon :};
				\node[right] at (-2.5,-5.5) { \(\lim_{u\to 0} u^2 g_{\mu\nu}^{Bulk}=g_{\mu\nu}^{boundary} \)  for \(\mu, \nu \neq u\)};
				\node[right] at (-9,0.5) {2. Bilinear
					combination :};
				\node[right] at (-9,-0.5) {\({g^{b}}^{\mu\nu}={e}^\mu_a{e}^\nu_b\eta^{ab}\)};
				\node[right] at (4.5,0.5) {4. Bilinear
					combination :};
				\node[right] at (4.5,-0.5) {\({g^{B}}^{\mu\nu}={E}^\mu_a{E}^\nu_b\eta^{ab}\)};
		\end{tikzpicture}}
		\caption{Diagram illustrating the process of relating holographic material to real material}	
	\end{figure}
	Through analytical calculations of the holographic Green's function, we derived the holographic fermion spectrum and confirmed that a Galilean boost produces a tilted Dirac cone structure within this holographic model. Furthermore, we computed the topological number of the system and verified that it remains invariant under tilting. This finding underscores the robustness of the system's topological properties against variations in tilt.
	
	Additionally, by deriving the effective metric of the material, we demonstrated that the feasibility of constructing an effective holographic model. This model enables a systematic study of the interplay between tilting and various physical properties, such as conductivity and symmetry-breaking effects.  Our  results   provide a novel framework for connecting holographic materials to real materials, paving the way for future explorations in the realm of strongly interacting tilted Dirac fermions.
	
	\bibliographystyle{JHEP}
	\bibliography{Refs.bib}

\end{document}